\documentclass[aip]{revtex4-2}

\usepackage{graphicx}        
\usepackage{color}           
\usepackage{siunitx}         
\usepackage{bm}              
\usepackage{makecell}        
\usepackage{hyperref}        
\usepackage{physics}

\begin{document}

\title{Quantum-coherent nonlinear interferometry using electron-phonon systems 
for entanglement-assisted terahertz sensing}
\author{Junya Ogiri}
\affiliation{Graduate School of Regional Development and Creativity, Utsunomiya University, 7-1-2 Yoto, Utsunomiya, Tochigi 321-8585, Japan}

\author{Hiroaki Minamide}
\affiliation{Tera-Photonics Research Team, RIKEN Center for Advanced Photonics, RIKEN,
Sendai, Miyagi 980-0845, Japan}

\author{Kunio Ishida}
\affiliation{School of Engineering, Utsunomiya University, 
7-1-2 Yoto, Utsunomiya, Tochigi 321-8585, Japan}
\email[]{ishd\_kn@cc.utsunomiya-u.ac.jp}

\begin{abstract}
  We develop a microscopic model of spontaneous parametric down-conversion (SPDC) in a nonlinear Mach-Zehnder interferometer mediated by an electron-phonon system.
  By treating the phonon mode as a coherent quantum degree of freedom, rather than as a dissipative reservoir, the model captures a two-stage buildup of entanglement and reveals phase-sensitive idler interference that encodes the internal light-matter coherence.
  These results provide a route to engineering entanglement as a resource in active quantum media with internal electron-phonon coherence and suggest relevance to quantum-enhanced sensing particularly in the terahertz regime.
\end{abstract}

\maketitle

\section{Introduction}
Recent advances in quantum information technology have brought renewed interest in quantum entanglement. 
In response, a wide range of physical systems have been explored for the generation of entangled states, including photons produced by biexciton cascades\cite{edmt} and entangled phonons in spatially separated systems\cite{lee,ki1}. 
Among these, spontaneous parametric down-conversion (SPDC), a second-order nonlinear optical process, has become a widely used source of entangled photon pairs\cite{bib6,bib7,bib8}, finding applications in quantum imaging\cite{image} and quantum-enhanced spectroscopy\cite{spec1,spec2,spec3}.
Efforts towards practical use of SPDC have spurred the development of advanced nonlinear optical materials, including microstructured inorganic crystals and organic ferroelectrics\cite{bib3,bib4}. 

Nonlinear interferometry is a key technique for harnessing entangled photons in quantum applications. 
Accordingly, precise control over the polarization, frequency, and wavevector of generated idler and signal photons has been achieved by phase-matching techniques, such as periodically poled structures\cite{bib5,bib9}.
Various nonlinear interferometer designs including the Zou-Wang-Mandel\cite{bib11} and SU(1,1) interferometers\cite{bib12} have been studied. 
These setups typically incorporate nonlinear crystals into an interferometric architecture such as a Mach-Zehnder or SU(1,1) geometry, and have been proposed as promising platforms for quantum metrology, particularly for sensitive detection in spectral regions (e.g., THz, far-infrared) where direct detectors are inefficient\cite{bib13}. 
Corresponding theoretical\cite{bib14,bib15,bib16} and experimental\cite{bib18,bib19} investigations have demonstrated phase-sensitive enhancement of detection sensitivity in those bands.

However, most previous studies treat nonlinear crystals merely as passive elements for photon conversion, neglecting the quantum coherence of the entire light-matter system. 
In conventional perturbative treatments, the material degrees of freedom appear only as virtual intermediate states in an effective nonlinear susceptibility $\chi^{(2)}$, and therefore the explicit quantum coherence of the coupled electron-phonon-photon system is missing.
This motivates a description where the medium is treated as an active quantum subsystem rather than a passive $\chi^{(2)}$ element.

To understand the entanglement structure between the input and output fields, it is essential to explicitly consider the quantum coherence within the nonlinear optical medium itself.
Therefore, the aim of this study is to investigate the dynamics of entangled photons generated via SPDC in a Mach-Zehnder-type interferometer, incorporating the quantum mechanical degrees of freedom associated with the material system.
The material systems are described by a spin-boson model that is used to discuss the dynamics of electron-phonon-photon systems\cite{ki2}.
Upon establishing the role of these internal degrees of freedom, we investigate the resulting quantum features in more detail.
In summary, we discuss the potential application of these systems for the quantum detection of weak terahertz waves.

We show below that explicitly retaining the electron-phonon-photon coherence reveals a two-stage buildup of entanglement and directly imprints this entanglement onto the interferometer output, enabling indirect readout of terahertz-band signal modes through near-infrared idler detection.

\section{Model}
\subsection{Electron-phonon-photon coupled system}
\begin{figure}[h]
\begin{center}
\scalebox{0.5}{\includegraphics{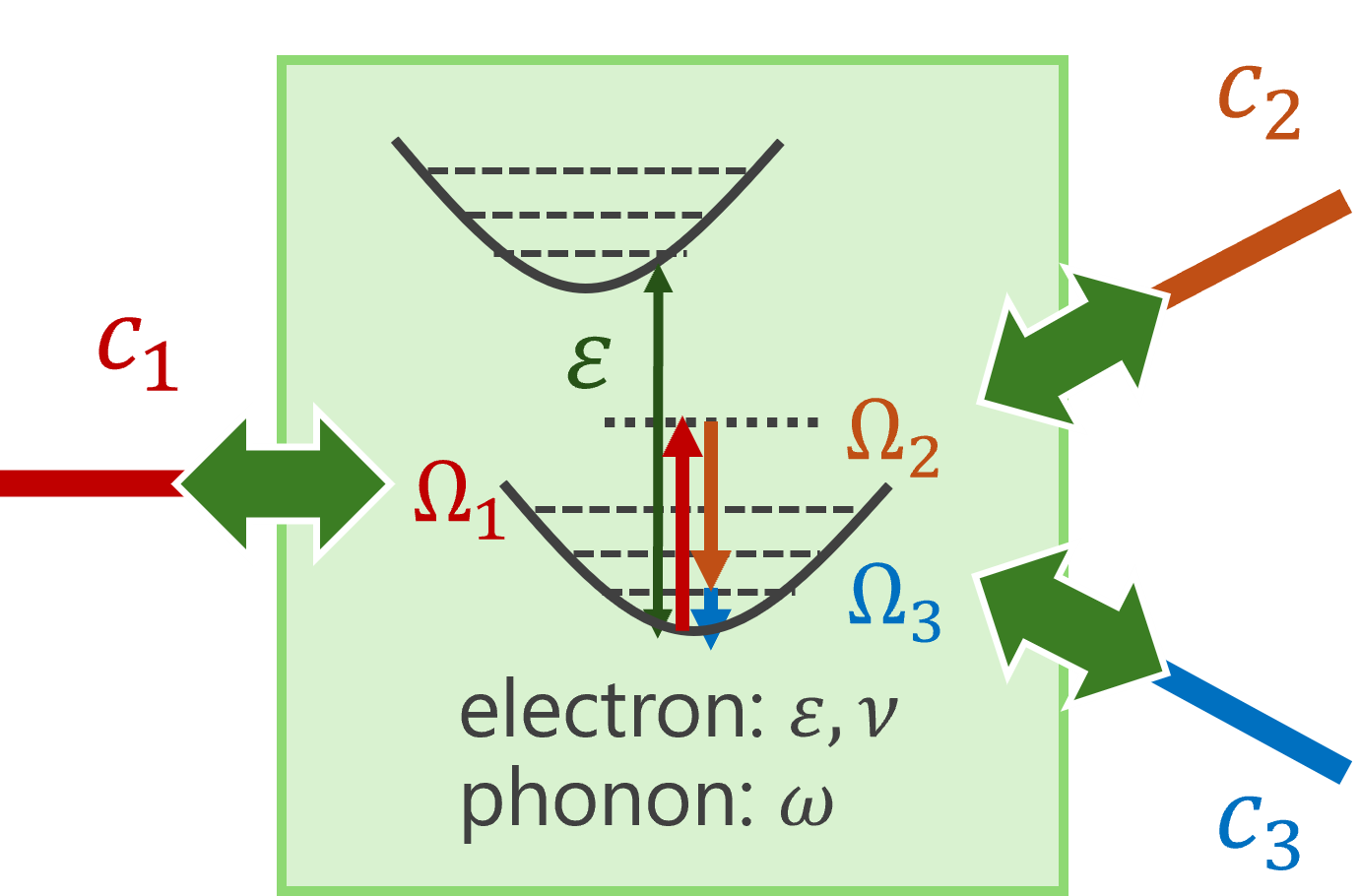}}
\caption{Schematic of electron-phonon-photon model used to describe SPDC process in the interferometer.}\label{model_fig}
\end{center}
\end{figure}

Since it is difficult to track the quantum dynamics of light-matter systems due to the large Hilbert-space dimension, we adopt a simplified model that still captures the essential physics to study the dynamics of quantum entanglement in SPDC-based interferometers.
We are particularly interested in operation in the terahertz regime, where the SPDC “signal” mode lies in the terahertz band. In this regime, phonons play an essential role in mediating the nonlinear process.
As a result, we employ a model of electron-phonon systems described by a spin-boson model in this paper.
As for the photon system, we consider three modes for pump, idler, and signal light, and the Hamiltonian of the entire electron-phonon-photon system is given by
\begin{equation}
\label{Hamiltonian_1mat}
{\cal H}_{0} = \Omega_{1} c^{\dag}_{1} c_{1} + \Omega_{2} c^{\dag}_{2} c_{2} + \Omega_{3} c^{\dag}_{3} c_{3} + \omega a^{\dag} a + \hat{n} \left\{ \nu \left( a^{\dag} + a \right) + \varepsilon \right\} + \hat{M} \sum_i (c^{\dag}_{i} + c_{i} ).
\end{equation}
where
\begin{eqnarray}
  \label{operator_population_1mat}
  \hat{n} &=& \frac{1 + \sigma_{z}}{2},\\
  \label{operator_dipole_1mat}
  \hat{M} &=&  \left\{ \mu + \tau \left( a^{\dag} + a \right) \right\} \sigma_{x} + \tau \left( a^{\dag} + a \right) ,
  \label{operator_electric_field_1mat}
\end{eqnarray}
and
$a$ and $c_i$ denote the annihilation operators of the optical phonons and the photons, respectively. The indices for $c_i$ correspond to the pump mode (1) , the idler mode (2), and the signal mode (3) with different frequencies, respectively.
We also set $\hbar = 1$.
$\sigma_{\rho}(\rho=x,y,z)$ are the Pauli matrices that operate on the two-level electronic states describing the ground state $|g\rangle$ and the excited state $|e \rangle$ of the electronic system. 
$\hat{M}$ is proportional to the transition electric dipole moment, and the last term of Eq.\ (\ref{Hamiltonian_1mat}) describes the light-matter interaction.
Since $M$ depends on the deformation of the lattice, $x$, we expand $M$ to the first order of $x$ ($\propto a^\dagger +a$), which is given up to the first order term of $x$ by
\begin{equation}
  \label{dipole_classic}
  M(x) = M(0) + \left( \frac{dM_i}{dx} \right)_{x=0}x + \cdots.
\end{equation}
This expansion shows that the SPDC process in our model is intrinsically accompanied by phonon creation/annihilation, since the signal photon frequency is comparable to the phonon frequency.
$\mu$ and $\tau$ represent the strength of the dipole interaction, and $\nu$ shows the strength of the electron-phonon interaction in the matter system. 
Figure \ref{model_fig} shows a schematic view of the SPDC process mediated by the current electron-phonon system.

\subsection{Equations of motion and numerical procedure}
\label{eomint}
To identify which microscopic processes actually generate and redistribute entanglement, we analyze the Heisenberg equations of motion for the photon annihilation operators and the coupled electron-phonon operators\cite{ki1,ki2}. 
\begin{equation}
  i \frac{d}{dt}\hat{O} = [\hat{O},\mathcal{H}],
  \label{Heisenberg_general}
\end{equation}
where $\hat{O}$ represents an arbitrary system operator.  
This formalism enables us to simultaneously trace the quantum dynamics of the photonic modes, the phonon mode, and the electronic operators in a self-consistent manner.

The general operator structure given by Eq.\ (\ref{Heisenberg_general}) forms the basis for the time-dependent analyses presented in Sec. \ref{results}.  
A detailed operator-level discussion focusing on which microscopic interactions predominantly contribute to entanglement generation is provided in Sec.\ \ref{eomresults}.

As for the values of the parameters, we set $\omega = 1$, $\mu = 0.5$, and $\tau = 0.1$.
The frequencies of the light fields were chosen as $\Omega_{1} = 13.5$, $\Omega_{2} = 12.5$, and $\Omega_{3} = 1$.
Throughout this study, time is expressed in units of $1/\omega$.
Since most ferroelectrics used as nonlinear optical crystals are insulating materials, we assume the energy band gap $\varepsilon = 27$, which shows that the present configuration corresponds to a nonresonant SPDC process.

All numerical calculations are performed in a truncated Fock basis with finite photon and phonon occupation cutoffs; we have verified that the results in Figs.\ \ref{observable_light_fig}-\ref{HEOM_fig} are converged with respect to these cutoffs.

\section{Nonlinear interferometer configurations}
This section introduces the nonlinear interferometer configurations considered in this study.
We first describe the two-stage architecture and its physical concept (Sec.\ \ref{tsni}), then formulate the corresponding Hamiltonians (Sec.\ \ref{hamil}), and finally define the detection and correlation measures (Sec.\ \ref{detect}).

\subsection{Two-stage nonlinear interferometer}
\label{tsni}
\begin{figure}[h]
\begin{center}
  \scalebox{0.4}{\includegraphics{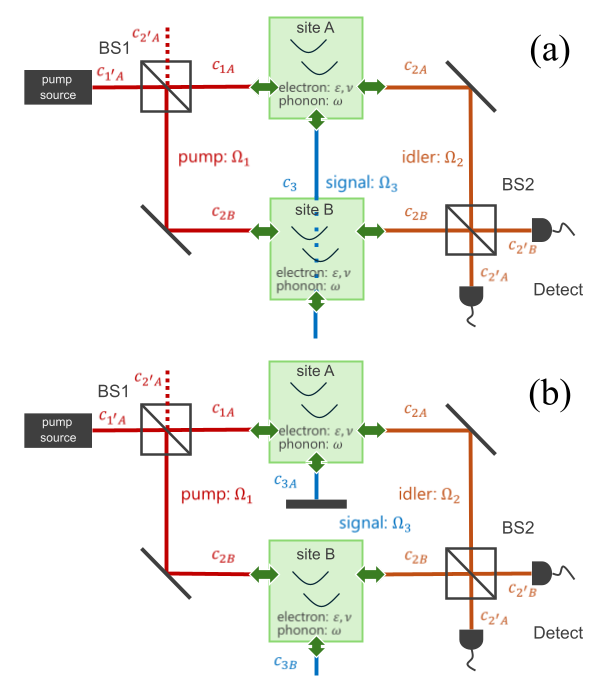}}
\caption{Schematic of the Mach-Zehnder-type interferometer with two identical electron-phonon coupled systems (A and B).
(a) Configuration C1: the signal mode generated in A is injected into B, so that both media share the same signal field. (b) Configuration C2: the signal path from A to B is blocked, so each medium interacts only with its local signal mode.}\label{model_int}
\end{center}
\end{figure}

We consider a Mach-Zehnder-type nonlinear interferometer composed of two identical electron-phonon coupled systems, as illustrated in Figure \ref{model_int}(a) and (b).
Each system acts as a nonlinear optical medium, where parametric interactions among electrons, phonons, and photons give rise to spontaneous parametric down-conversion (SPDC).
The two configurations differ in the connection topology of the signal path between the systems.
In configuration C1, the signal mode $c_3$ generated in system A is injected into system B, so that both systems share the same signal field.
In configuration C2, this coupling path is blocked, and each system interacts only with its local signal mode.
In C1, the idler and signal photons generated in A are entangled through the local quantum coherence within A, and this entanglement is subsequently transferred to B via the shared signal field.
As a result, the idler photons emitted from A and B become mutually entangled in C1.
In contrast, in C2 the signal photons do not propagate into B, and no nonlocal entanglement between the two idler beams is formed.
The essential idea of this study is to use the entanglement properties between the two idler outputs as a probe for coherence dynamics of the signal photons.

\subsection{Configurations C1 and C2}
\label{hamil}

We introduce Hamiltonians for the two configurations C1 and C2, corresponding to the setups illustrated in Figure \ref{model_int}(a) and (b).

For configuration C1, where the signal mode generated in system A is injected into system B, the Hamiltonian is written as
\begin{equation}
\label{Hamiltonian_1}
{\cal H} _1= \Omega_{3} c^{\dag}_{3} c_{3} + \sum_{j = A, B} \left[ \Omega_{1} c^{\dag}_{1j} c_{1j} + \Omega_{2} c^{\dag}_{2j} c_{2j} + \omega a^{\dag}_{j} a_{j} + \hat{n}_{j} \left\{ \nu \left( a^{\dag}_{j} + a_{j} \right) + \varepsilon \right\} + \hat{M}_{j} \hat{A}_{j} \right],
\end{equation}
where
\begin{eqnarray}
  \label{operator_population}
  \hat{n}_{j} &= &\frac{1 + \sigma^{j}_{z}}{2},\\
  \label{operator_dipole}
  \hat{M}_{j} &= &\left\{ \mu + \tau \left( a^{\dag}_{j} + a_{j} \right) \right\} \sigma^{j}_{x} + \tau \left( a^{\dag}_{j} + a_{j} \right),
\end{eqnarray}
and 
\begin{equation}
  \label{operator_electric_field}
  \hat{A}_{j} =  \left  ( c^{\dag}_{1j} +  c^{\dag}_{2j} +  c^{\dag}_{3} \right )+  h.c.
\end{equation}
Here, the subscript $j = A, B$ denotes the two distinct matter systems.

For configuration C2 shown in Fig.\ \ref{model_int}(b), the signal photons from system A do not propagate to B, and the two systems operate independently.
The corresponding Hamiltonian is
\begin{equation}
\label{Hamiltonian_2}
{\cal H}_2 =  \sum_{j = A, B} \left[ \Omega_{1} c^{\dag}_{1j} c_{1j} + \Omega_{2} c^{\dag}_{2j} c_{2j} + \Omega_{3} c^{\dag}_{3j} c_{3j}+ \omega a^{\dag}_{j} a_{j} + \hat{n}_{j} \left\{ \nu \left( a^{\dag}_{j} + a_{j} \right) + \varepsilon \right\} + \hat{M}_{j} \hat{A}'_{j} \right],
\end{equation}
where
\begin{equation}
\hat{A}^{\prime}_{j} = \left( c^{\dag}_{1j}  + c^{\dag}_{2j} + c^{\dag}_{3j}  \right) + h.c.
\end{equation}
In the following, the above two setups are referred to as C1 and C2, respectively.

\subsection{Detection and phase readout}
\label{detect}
The beam splitters BS1 and BS2 used in the interferometer are treated using the standard quantum beam splitter formalism\cite{bib22}.
The input and output field operators satisfy the following relations:
\begin{eqnarray}
  \label{BS_relation}
  \left ( \begin{array}{cc}
  c_{1A} & c_{1B}
  \end{array}
\right )
 & = & \frac{1}{\sqrt{2}} \left (\begin{array}{cc}  1 & i \\ i & 1 \end{array} \right )
  \left ( \begin{array}{cc} c_{1^{\prime}A} &  c_{1^{\prime}B} \end{array} \right ),\\
  \left ( \begin{array}{cc}
  c_{2'A} & c_{2'B}
  \end{array}
\right )
 & = & \frac{1}{\sqrt{2}} \left (\begin{array}{cc}  1 & i \\ i & 1 \end{array} \right )
  \left ( \begin{array}{cc} c_{2A} &  c_{2B} \end{array} \right ).
\end{eqnarray}

The pump beam input is prepared in a coherent state $|\alpha\rangle$ with $\alpha=\sqrt{32}$, corresponding to an average photon number of 32 in the input arm.
All idler and signal beams are initially in the vacuum state, and the initial system state is
\begin{equation}
  |\Phi(0) \rangle = |\alpha/\sqrt{2}\rangle_{1A} \otimes |i\alpha/\sqrt{2}\rangle_{1B} \otimes |0\rangle_{2A} \otimes |0 \rangle_{2B} \otimes | 0 \rangle_{3} \otimes |G \rangle_{A} \otimes |G \rangle_{B},
  \end{equation}
where $|G \rangle_{j}$ denotes the ground state of the electron-phonon system at site $j$, which indicates that no quantum entanglement among the light and matter subsystems at $t = 0$.

To quantitatively analyze the entanglement dynamics of the system, it is essential to adopt an appropriate measure of quantum entanglement. 
Since the system comprises multiple interacting subsystems, {\it i.e.}, electronic, phononic, and photonic degrees of freedom, we employ quantum mutual information as a suitable metric, following the approach used in our previous studies\cite{ki1,ki2}.
The quantum mutual information is defined by
\begin{equation}
  I_{M} = S_{i_A} + S_{i_B} - S_{i_A \otimes i_B},
\end{equation}
where $S_{\alpha}$ is the von Neumann entropy for subsystem $\alpha$.
The calculation of  $I_M$ is detailed in the Supplementary Material\cite{suppl}.
This detection scheme and correlation measure serve as the foundation for the time-dependent analyses presented in Section \ref{results}.
The relative phase between the two idler paths is controlled before BS2 and subsequently read out via the output statistics at BS2 (Sec.\ \ref{p_sensitive}).

\section{Results}
\label{results}

\subsection{SPDC process in the model}

\begin{figure}[h]
\begin{center}
\scalebox{0.7}{\includegraphics{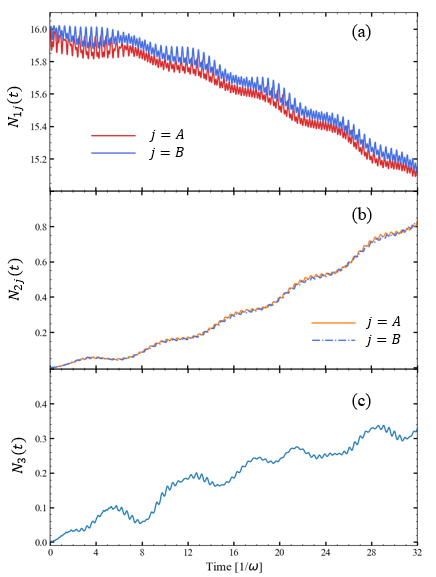}}
\caption{(a) The average number of pump photons $N_{1j}(t)= \ev{c^{\dagger}_{1j}c_{1j}}$ for $j=A, B$, (b) The average number of idler photons $N_{2j}(t) = \ev{c^{\dagger}_{2j}c_{2j}}$ for $j=A, B$, (c) The average number of signal photons $N_{3}(t) = \ev{c^{\dagger}_{3}c_{3}}$}
\label{observable_light_fig}
\end{center}
\end{figure}

We first examine the photon dynamics by tracking the average photon numbers of the pump $N_p$, idler $N_{i\alpha}$ $(\alpha=A,B)$, and signal $N_s$ light, and the calculated results are shown in Figs.\ \ref{observable_light_fig}(a)-(c) as functions of time.
These quantities show relatively slow variations over time, superimposed on faster oscillations at the phonon frequency (period $2\pi/\omega$). 
The number of pump photons decreases by about one at $t=32$, while the idler and signal photon numbers increase by approximately 0.8 and 0.4 in the same time interval. 
This behavior reflects the SPDC process accompanied by the quantum coherence of the matter system.
This already indicates that simple photon-number measurements are insufficient to distinguish the two configurations.
Although the idler intensities are almost identical in C1 and C2, we will show in the next subsection that their entanglement structure is dramatically different.

We also note that, when the signal photons from matter A propagate to matter B in C1, difference-frequency generation (DFG) in B is expected to occur in principle.
However, our calculations show that DFG contributions are negligible due to the small number of signal photons.

We find that the idler and signal photons are generated through different pathways.
Idler photons predominantly arise from electronic transitions driven by $\sigma_x$, while signal photon generation involves phonon creation. Consequently, the idler and signal photon dynamics differ, and no one-to-one correspondence exists between a pump photon and an idler-signal photon pair.

\subsection{Buildup of nonlocal entanglement}
\begin{figure}[h]
\begin{center}
  \scalebox{0.5}{\includegraphics{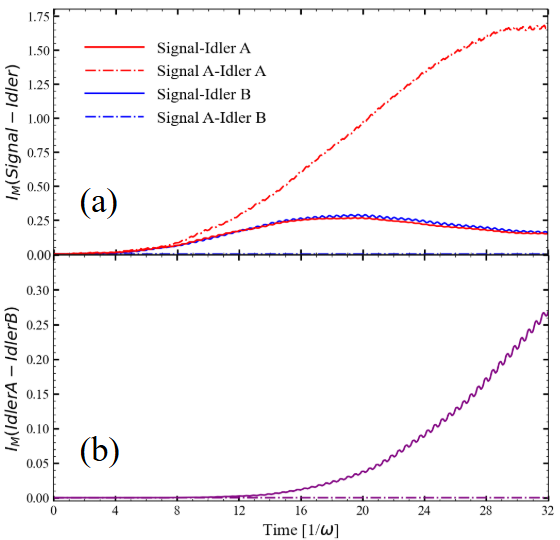}}
  \caption{Quantum mutual information $I_{M}(t)$. (a) Signal-idler $j$ $(j=A,B)$. (b) Idler $A$-idler $B$. The solid and dash-dotted lines correspond to those for C1 and C2, respectively.}
  \label{IM_photon_fig}
\end{center}
\end{figure}

When we focus on the entanglement dynamics, the quantum effect in the current configurations becomes apparent.
Figure \ref{IM_photon_fig}(a) shows the quantum mutual information between the idler photons and the signal photons, where the solid and dash-dotted lines correspond to those for C1 and C2, respectively.
The red and the blue lines correspond to $I_M(\text{signal, idler}\: A)$ and $I_M(\text{signal, idler}\: B)$, respectively.
For C1, both properties gradually increase after the pump irradiation begins, and are almost identical to each other.
We note that the entanglement between the idler photons and the signal photons is mediated by the quantum coherence of the electron-phonon systems.
In C2, the signal is blocked before entering B, and $I_M(\text{signal, idler}\: B)$ vanishes.
By contrast, $I_M(\text{signal, idler}\: A)$ increases, which is reminiscent of the entanglement monogamy\cite{monogamy}, i.e., the total information is distributed between $I_M(\text{signal, idler}\: A)$ and $I_M(\text{signal, idler}\: B)$.
Hence, the difference between C1 and C2 represents the role of signal photons in distributing entanglement across the interferometer.

As we showed in Ref.\ \cite{ki1}, the quantum nature of light induces entanglement between remote systems.
In a similar manner, the signal light mediates entanglement between two matter systems A and B in the present interferometric configuration.
As a result, the entanglement between idler photons from A and B behaves differently in C1 and C2.
Figure \ref{IM_photon_fig}(b) shows the quantum mutual information between the idler photons from $A$ and the idler photons from $B$ ($I_{M}(\text{idler A,  idler B})$), where the solid and dash-dotted lines correspond to those for C1 and C2, respectively.

In  C1, $I_M(\text{idler A, idler B})$ increases only after $t \sim 10 $, indicating a delayed onset of idler-idler entanglement that is mediated by the shared signal mode and the internal electron-phonon coherence.
In contrast, in C2 , where the signal from A is blocked,  this idler-idler entanglement is essentially absent.
More precisely, the idler photons from A and B are entangled with each other through the quantum coherence of the two matter systems (A and B) and the signal photons, and thus it is necessary to take into account their quantum nature for quantitative discussion on the photoinduced entanglement in C1 and/or C2.
Since the idler-idler entanglement is mediated not only by signal photons but also by electrons and phonons, its dynamics is influenced by a complex interplay of interactions, which is discussed in Sec.\ \ref{eomresults}.
Overall, the photon number alone does not necessarily reflect the quantum properties of light.

\subsection{Phase-sensitive idler interference}
\label{p_sensitive}
\begin{figure}[h]
\begin{center}
  \scalebox{0.3}{\includegraphics{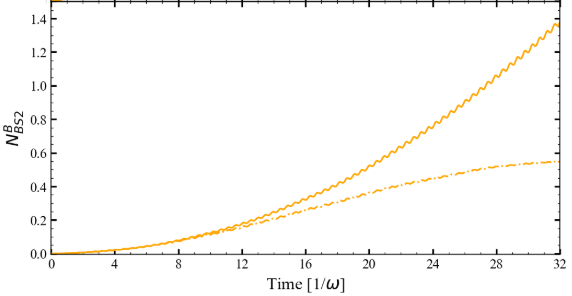}}
\caption{Output idler photon number from BS2, $N_{BS2}=\langle c_{2'A}^\dagger c_{2'A} \rangle$. The solid and dash-dotted lines correspond to C1 and C2, respectively.}
\label{bs2output}
\end{center}
\end{figure}

The present results show that this indirect access to the signal field becomes explicit when we calculate the output photon number at BS2, $N_{\mathrm{BS2}} = \langle c_{2'A}^\dagger c_{2'A}\rangle$, shown in Fig.\ \ref{bs2output}.
We find that, although $N_{BS2}$ remains finite for both C1 and C2, a remarkable difference is observed between the two configurations.
Since the idler photon number from A and B is nearly equal, we attribute this difference to phase-sensitive idler-idler entanglement, which modifies the output statistics of BS2.
This highlights that the BS2 output is sensitive to idler-idler entanglement and suggests indirect detection of signal photons through idler photon correlations. 
Importantly, if the signal mode lies in the terahertz band while the idler lies in the near-infrared, this mechanism provides an indirect, entanglement-assisted pathway for terahertz detection without requiring high-efficiency terahertz photon counting.

\subsection{Analysis of entanglement generation mechanism}
\label{eomresults}

\begin{figure}[h]
\begin{center}
\scalebox{0.8}{\includegraphics{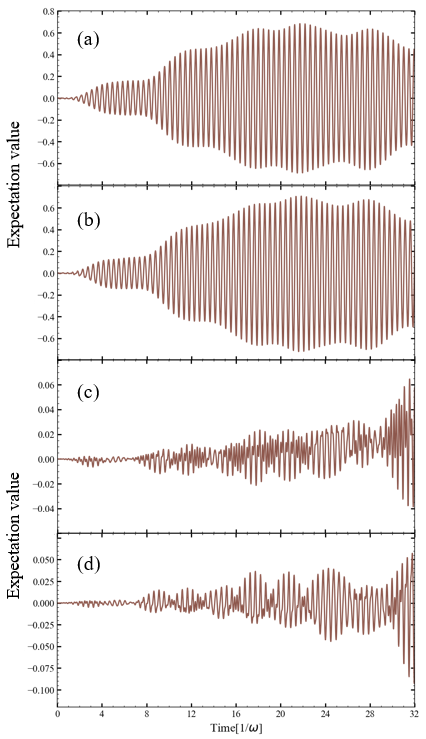}}
\caption{Expectation values of representative higher-order composite operators obtained from the Heisenberg equations of motion. (a) $\ev{\sigma^{A}_{z}c_{3}c_{2A}+\text{h.c.}}$, (b) $\ev{\sigma^{B}_{z}c_{3}c_{2B}+\text{h.c.}}$, (c) $\ev{\sigma^{A}_{y}c_{3}c_{2A}c^{\dag}_{2B}+\text{h.c.}}$, (d) $\ev{\sigma^{B}_{z}\sigma^{A}_{x}c_{3}c_{2A}c^{\dag}_{2B}+\text{h.c.}}$ }
\label{HEOM_fig}
\end{center}
\end{figure}

To elucidate the microscopic origin of the entanglement behaviors observed in Figs.\ \ref{observable_light_fig}-\ref{bs2output}, we analyze the operator dynamics based on the Heisenberg equations of motion introduced in Sec. \ref{eomint}.  
The equation of motion for $c_{2j}$ is given by
\begin{equation}
\label{HEOM_0_order_ex}
  i \frac{d}{dt} c_{2j}(t) = \Omega_{2} c_{2j}(t) + \mu \sigma_{x}^{j}(t) + \tau \left( a_{j}^{\dag}(t) + a_{j}(t) \right) \sigma_{x}^{j}(t)
  + \tau \left( a^{\dag}_{j}(t) + a_{j}(t) \right).
\end{equation}
The first term corresponds to the free evolution of idler photons, and the other terms show the effects of the light-matter interactions.
Then, we obtain the equation of motion for $\mu \sigma_{x}^{j}(t)$ that appears in Eq.\ (\ref{HEOM_0_order_ex}), for example, as
\begin{equation}
\label{HEOM_1_order_ex}
    i \dv{t} \mu \sigma_{x}^{j}(t) = - \mu \nu i \sigma_{y}^{j}(t) \left( a_{j}^{\dag}(t) + a_{j}(t) \right) - \mu \varepsilon i \sigma_{y}^{j}(t).
\end{equation}
This equation shows that $\sigma_{x}^{j}(t)$ couples to additional composite operators such as $\sigma_{y}^{j}(t) a_{j}^{\dag}(t)$ and $\sigma_{y}^{j}(t)$, where the latter arises from electron-phonon interactions.
In a similar manner various composite operators involving electrons, phonons, and photons are obtained order by order due to the interaction between pump photons, idler photons, signal photons, electrons, and phonons. 
Although the resulting hierarchy of operator equations does not close at any finite order, we compute them iteratively and evaluate the corresponding expectation values numerically.

Figure \ref{HEOM_fig}(a) and (b) show the expectation values of the third-order composite operators related to signal - idler $j$ entanglement, and Figs.\ \ref{HEOM_fig}(c) and (d) show those of the sixth-order operators related to idler A-idler B entanglement.
We note that these operators do not appear at any order of the equations of motion for C2.
Since the Hamiltonian is Hermitian, each equations of motion has a Hermitian-conjugate counterpart and it is sufficient to calculate their real parts.
Figure \ref{HEOM_fig}(a) and (b) show that the contribution of $\ev{\sigma^{A}_{z}c_{3}c_{2A}}$ and $\ev{\sigma^{B}_{z}c_{3}c_{2B}}$ to the dynamics is negligible for $0 < t < 2$ compared to those for $t > 16$, whereas Figs.\ \ref{HEOM_fig}(c) and (d) show that the values of $\ev{\sigma^{A}_{y}c_{3}c_{2A}c^{\dag}_{2B}+\text{h.c.}}$ and $\ev{\sigma^{B}_{z}\sigma^{A}_{x}c_{3}c_{2A}c^{\dag}_{2B}}$ stay small for $0 < t < 8$.
These results confirm the delayed onset of quantum entanglement after pump irradiation, consistent with the dynamical behavior of $I_M$ in Fig.\ \ref{IM_photon_fig}.
In this way, the dynamics of composite operators reflect the cooperative entanglement generation mechanism of the coupled light-matter system.

\subsection{Implications for quantum sensing}
While our model is theoretical, its implementation can be realized experimentally with current photonic technologies 
including shared terahertz waveguides or free-space terahertz coupling between materials A and B. 
Although phase stability and decoherence mitigation during terahertz propagation are critical, recent advances in terahertz photonics make such configurations feasible.
The use of BS2 for idler recombination does not require terahertz manipulation, making this scheme especially attractive for sensing applications where sensitive terahertz detection is otherwise difficult.

Our approach differs substantially from conventional nonlinear interferometry schemes, such as SU(1,1) interferometers or cascaded $\chi^{(2)}$ processes in optical parametric oscillators. Table \ref{tab:comparison} summarizes the key distinctions.

\begin{table}[t]
\caption{Side-by-side comparison of schemes.\footnote{THz denotes feasibility of indirect terahertz detection via idler correlations.}}
\centering
\scriptsize
\setlength{\tabcolsep}{4pt}\renewcommand{\arraystretch}{1.15}
\begin{tabular*}{\linewidth}{@{\extracolsep{\fill}} l l l l}
Scheme & Role of nonlinear medium & Entanglement pathway & THz sensing via idler readout\\
\hline
$\chi^{(2)}$ SPDC (std.) & Passive medium & Classical polarization response & No \\
SU(1,1) interferometer & Amplifying crystals & Gain-mediated entanglement & Indirect/bandwidth-limited \\
This work & Quantum-coherent matter & \makecell[l]{Internal electronic/phononic coherence; \\ idler-idler correlations} & Yes \\
\hline
\end{tabular*}
\label{tab:comparison}
\end{table}

Our model shows that the role of the nonlinear medium should be regarded as an active quantum system that facilitates entanglement through its internal degrees of freedom rather than a passive converter.
This distinction not only enriches the theoretical framework but also opens up new sensing technologies unavailable to traditional interferometric schemes.

\section{Conclusions}
We study the dynamics of entangled photon generation by SPDC via electron-phonon systems embedded in a Mach-Zehnder interferometer and its effect on the output photons.
By considering the quantum coherence of the nonlinear optical materials, we obtained the quantum entanglement dynamics of the entire electron-phonon-photon system.

Our calculations reveal two-stage entanglement generation: first between signal and idler photons, followed by delayed idler-idler entanglement mediated by the quantum coherence of phonons and electrons. 
An analysis by the Heisenberg equations of motion reveals the cooperative light-matter nature of this process.
The output of BS2, modulated by these entanglement dynamics, suggests a potential quantum-enhanced detection scheme for weak terahertz waves.
Beyond conventional nonlinear susceptibility $\chi^{(2)}$ descriptions, our model clarifies the significance of time-dependent quantum coherence within the material systems. 

This architecture can be viewed as a hybrid interferometer in which the nonlinear elements are themselves quantum systems with internal electron-phonon-photon coherence, rather than passive $\chi^{(2)}$ media.
Future extensions along this direction will incorporate spatial mode structure, loss, and decoherence, and will benchmark the present mechanism against realistic experimental parameters.
In this work we focus on coherent unitary dynamics and neglect propagation loss and environmental decoherence, in order to isolate the intrinsic buildup of light-matter entanglement.
Incorporating realistic loss channels will be essential for quantitative sensitivity estimates in an experimental setting.

\section*{Acknowledgments}

The authors are grateful to D. Yadav and N. Aoyagi for fruitful discussions.
Numerical calculations were performed on the facilities of the Supercomputer Center, Institute for Solid State Physics, the University of Tokyo, Japan.
This work was supported by JSPS KAKENHI Grant Number JP25K08502.

The data that support the findings of this study are available from the corresponding author upon reasonable request.

\end{document}